# APERIODIC ISING MODELS

UWE GRIMM
*Institut voor Theoretische Fysica, Universiteit van Amsterdam,*
*Valckenierstraat 65, 1018 XE Amsterdam, The Netherlands*

AND

MICHAEL BAAKE
*Institut für Theoretische Physik, Universität Tübingen,*
*Auf der Morgenstelle 14, 72076 Tübingen, Germany*

**Abstract.** We consider several aspects of non-periodic Ising models in one and two dimensions. Here we are not interested in random systems, but rather in models with intrinsic long-range aperiodic order. The most prominent examples in one dimension are sequences generated by substitution rules on a finite alphabet. The classical one-dimensional Ising chain as well as the Ising quantum chain with coupling constants modulated according to substitution sequences are considered. Both the distribution of Lee-Yang zeros on the unit circle in the classical case and that of fermion frequencies in the quantum model show characteristic gap structures which follow the gap labeling theorem of Bellissard. We also investigate the zero-field Ising model on two-dimensional aperiodic graphs, which are constructed from rectangular grids in the same spirit as the so-called Labyrinth. Here, duality arguments and exact solutions in the sense of commuting transfer matrices are used to gain information about the critical behaviour.

## 1. Introduction

The study of lattice spin models has long been a major subject of statistical physics. These systems, of which the Ising model is among the simplest, are interesting both from a mathematical point of view and for applications in physics. In particular, the universality hypothesis states that the behaviour of such systems close to critical points (provided the interactions are of finite range) depends only on very general properties as, for instance, the dimension and symmetries of the model. This means that real systems in the vicinity of second-order phase transitions (for example magnets at the Curie temperature) may well be described by very simple toy models which only capture the essential ingredients.

Unfortunately, even such a simple model as the Ising model has not been solved in three dimensions, which clearly is the physically most interesting case. However, 1D and 2D models not only initiated important new developments in mathematics and theoretical physics, but also found various physical applications (Stinchcombe, 1973; Einstein, 1987).

While there is a wealth of detail and rigorous treatment of periodic and related systems (Ruelle, 1969; Ellis, 1985; Simon, 1993), the knowledge about random systems is still rather vague and that about aperiodic systems with long-range orientational order hardly existing. One exception is the article by Geerse and Hof (1991), where an approach to equilibrium states for lattice gas models on self-similar tilings is made, but, apart from that, few exact results about the periodic case have found an aperiodic counterpart.

In this article, we present results on one- and two-dimensional aperiodic Ising models, where the aperiodicity is implemented by means of sequences based on substitution rules. We should emphasize that we do not aim at giving final answers to the problems connected with aperiodicity in lattice models, but rather collect useful insight for further, and more rigorous, investigations.

The article is organized as follows. After brief remarks about the history of the Ising model, Section 2 gives a short review of the properties of (periodic) Ising models in one and two dimensions. At the same time, this part introduces our notation and some basic concepts. In Section 3 we introduce the classical 1D Ising model (in constant magnetic field) with an aperiodic sequence of coupling constants. We investigate its Lee-Yang or magnetic field zeros which (in proper variable) lie on the unit circle but show some unexpected behaviour: generically they only fill a Cantor subset of it (a first step of a proof is given in the Appendix).

Next, Section 4 deals with its quantum mechanical counterpart, the Ising quantum chain with transversal field. It corresponds to an anisotropic limit of the the classical 2D Ising model without magnetic field (Fradkin and Susskind, 1978; Kogut, 1979). Starting from the Harris (1974) criterion for relevant disorder and Luck's (1993) adaption to quantum chains, we discuss a selection of typical examples with respect to critical behaviour and conformal invariance. It turns out that the latter is lost if fluctuations in the couplings are relevant.

In Section 5 we move on to classical 2D Ising models (in zero external field) on non-periodic graphs. By means of Baxter's (1978) concept of "$Z$-invariance" we derive an exact solution for a subclass of models with a phase transition of the Onsager universality class. In contrast to the situation in Section 4, this result is robust to a larger class of fluctuations.

Though this article is not a review, we have spent some effort to include a rather complete list of articles (not all of which are cited in the text) in the Bibliography at the end of this paper.

## 2. The Ising Model

The Ising model is, without doubt, the most famous and best understood among the models of statistical mechanics. It was proposed by Lenz (1920) as a toy model of ferromagnetism. Five years later, his student Ising (who recently celebrated his 95th birthday) published the solution of the 1D model (Ising, 1925), which showed no phase transition at finite (i.e., non-zero) temperature, and Ising supposed the same to be true in any dimension.

This conclusion, however, proved wrong. In fact, Peierls (1936) showed that in two and more dimensions the Ising model undergoes a phase transition from a disordered high-temperature to an ordered low-temperature phase. By duality arguments, Kramers and Wannier (1941) located the critical point of the 2D zero-field square lattice Ising model, assuming that the phase transition is unique. Shortly after, Onsager (1944) published his famous solution for the partition function of this model. The spontaneous magnetization in the ordered phase and its power law behaviour with the critical exponent $\beta = 1/8$ was announced by Onsager and calculated soon after (Yang, 1952; Montroll et al., 1963). Correlation functions have also been studied, see for instance (McCoy and Wu, 1973). For further details on the history and an extensive bibliography we refer to (Brush, 1967).

Today, inspired by Onsager's work (Baxter, 1995), a plethora of so-called "solvable models" in two dimensions is known (Jimbo, 1989). These are constructed as solutions of the *Yang-Baxter equation* (YBE) which basically is Onsager's *star-triangle relation* (see (Au-Yang and Perk, 1989) for its history). The YBE implies that row transfer matrices (Baxter, 1982) form a one-parameter family of commuting matrices. In a sense, the corresponding models can be regarded as generalizations of the Ising model, but for most of them solutions are only known (under a list of assumptions) in the thermodynamic limit whereas Onsager was also able to compute the complete finite-size spectrum of the Ising model.

To the present day, the 2D Ising model in a magnetic field has not been solved. However, some properties of this system are known from conformal field theory (Zamolodchikov, 1989). Moreover, a solvable model has recently been found (Warnaar et al., 1992) that (in the sense of universality) describes the same physics as the Ising model at the critical temperature in the presence of a symmetry-breaking field.

By introducing disorder (for instance by considering Ising models with random coupling constants) into a model one might change its physical properties considerably, see (Koukiou, 1993; McCoy, 1995) and references therein. There are general qualitative arguments about the effects of disorder. Clearly, one would expect that critical singularities, if affected at all, can only be "softened" by randomness because these are co-operative phenomena. This is precisely what was observed by McCoy and Wu (1968): a certain kind of layered disorder destroys the logarithmic singularity of the specific heat at the phase transition of the 2D Ising model. A well-known criterion for the relevance of disorder to critical properties is due to Harris (1974) which depends on the sign of the specific heat exponent $\alpha$ in the unperturbed model: critical singularities are smoothed out for $\alpha > 0$, whereas disorder is irrelevant for $\alpha < 0$. However, since a logarithmic divergence of the specific heat means $\alpha = 0$, this criterion does not give a definite answer for the 2D Ising model. Numerous investigations of random bond Ising models using analytical and numerical techniques have not come up with a definite answer either, but in many cases no differences between the random and periodic case could be detected.

In what follows, we are interested in a kind of aperiodicity which is different from randomness, and in fact constitutes a long-range order in the system (wherefore the term "disorder" is not appropriate). Typical examples are Ising models with coupling constants chosen according to (completely deterministic) substitution sequences, or Ising models on quasiperiodic graphs such as the Penrose tiling. In particular for the latter there is an extensive literature of mainly numerical results, mostly based on Monte Carlo simulations. To make some progress with open questions, we focus on exact results. We now set the scene by briefly reviewing the periodic case.

## 2.1. 1D ISING CHAIN

Consider a linear array of $N$ "Ising spin" variables $\sigma_j$, enumerated by $j = 1, \ldots, N$ (see Figure 1), which take values $\sigma_j \in \{1, -1\}$ (or "+" and "−", for short). To any *configuration* $\boldsymbol{\sigma} = \{\sigma_j \mid j = 1, \ldots, N\} \in \{1, -1\}^N$ a quantity (*energy*) is assigned through

$$E(\boldsymbol{\sigma}) = -\sum_{j=1}^{N} J_j \sigma_j \sigma_{j+1} - \sum_{j=1}^{N} H_j \sigma_j \tag{1}$$

where it is convenient to use *periodic boundary conditions*, i.e., $\sigma_{N+1} = \sigma_1$. For the moment, we keep our notation general and allow the *coupling constants* $J_j$ and the local values of the *magnetic field* $H_j$ to depend on the position in the chain. The model is called *ferromagnetic* if the energy weight favours the variables on neighbouring sites to coincide, i.e., if the coupling constants $J_j$ are positive. Although many results remain valid for the *antiferromagnet* as well, we will usually assume that all couplings are ferromagnetic. In particular, we do not discuss Ising models with a mixture of both ferro- and antiferromagnetic couplings which already in one dimension show complicated *frustration* effects, see e.g. (Luck, 1987).

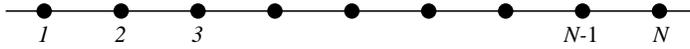

**Figure 1.** One-dimensional Ising chain

The *statistical* or *Boltzmann weight* of a configuration $\boldsymbol{\sigma}$ is

$$W(\boldsymbol{\sigma}) = \prod_{j=1}^{N} W_j(\sigma_j, \sigma_{j+1}) = \exp(-\beta E(\boldsymbol{\sigma})) \tag{2}$$

which is a product of *local Boltzmann weights*

$$W_j(\sigma, \sigma') = \exp\left(K_j \sigma \sigma' + \frac{1}{2}(h_j \sigma + h_{j+1}\sigma')\right) \tag{3}$$

associated to the bond between sites $j$ and $j+1$. Here, we introduced dimensionless quantities $K_j = \beta J_j$ and $h_j = \beta H_j$, where $\beta = 1/k_B T$ with temperature $T$ and Boltzmann's constant $k_B$.

The (canonical) *partition function* is defined by

$$Z_N = \sum_{\boldsymbol{\sigma}} \exp\left(-\beta E(\boldsymbol{\sigma})\right) \tag{4}$$

where the sum is over all $2^N$ configurations $\boldsymbol{\sigma}$. Then $W(\boldsymbol{\sigma})/Z_N$ defines a probability measure (a finite volume *Gibbs state*) and expectation values (*correlation functions*) are computed with respect to this measure (for example the *local magnetization* $\langle \sigma_j \rangle = \sum_{\boldsymbol{\sigma}} \sigma_j W(\boldsymbol{\sigma})/Z_N$ at site $j$). Introducing local *transfer matrices* $\boldsymbol{T}_j$ defined by

$$\boldsymbol{T}_j = \begin{pmatrix} W_j(+,+) & W_j(+,-) \\ W_j(-,+) & W_j(-,-) \end{pmatrix} = \begin{pmatrix} e^{K_j + (h_j + h_{j+1})/2} & e^{-K_j + (h_j - h_{j+1})/2} \\ e^{-K_j - (h_j - h_{j+1})/2} & e^{K_j - (h_j + h_{j+1})/2} \end{pmatrix}$$

allows us to rewrite the sum over configurations in the partition function in terms of matrix multiplications

$$Z_N = \sum_{\boldsymbol{\sigma}} \prod_{j=1}^{N} W_j(\sigma_j, \sigma_{j+1}) = \mathrm{tr}(\boldsymbol{T}_1 \boldsymbol{T}_2 \cdots \boldsymbol{T}_N) = \mathrm{tr}(\mathcal{T}) \tag{5}$$

where the trace is due to periodic boundary conditions. The matrix $\mathcal{T}$ is frequently referred to as *monodromy matrix* in the literature. The (specific) *free energy* is given by

$$f_N = -\frac{1}{\beta N} \log(Z_N) \tag{6}$$

from which other thermodynamic quantities result as suitable derivatives.

One is particularly interested in the *thermodynamic limit* in which the system size becomes infinite, and especially in points where the resulting free energy develops non-analytic behaviour as a function of temperature which can be regarded as a definition of a *phase transition* (Griffiths, 1972; Thompson, 1979). Clearly, if the local weights are analytic functions of temperature (as is usually the case), phase transitions can only appear in the thermodynamic limit. Of particular interest are so-called *critical points* (or second-order phase transitions) which are those phase transition points where the first derivative of the free energy is still continuous, while the second is not. These are characterized by correlation functions which show algebraic (with powers that are known as *critical exponents* or *critical indices*) rather than the usual exponential decay (with characteristic distances called *correlation lengths*). In other words, correlation lengths diverge if one approaches a critical point, which means that fluctuations (not just microscopically, but at all length scales) become important and critical systems can effectively be described by continuum field theories, see Section 2.4.

In general, it is impossible to calculate the partition function (5) explicitly, even for this simple 1D model. However, if all the transfer matrices $\boldsymbol{T}_j$ commute with each other, we can diagonalize them simultaneously. For example, for the translationally invariant case $J_j \equiv J$ and $H_j \equiv H$, one obtains $\boldsymbol{T}_j \equiv \boldsymbol{T}$ for all $j$ (hence $\mathcal{T} = \boldsymbol{T}^N$) and

$$f_N = -\frac{1}{\beta N} \log\left(\mathrm{tr}(\boldsymbol{T}^N)\right) = -\frac{1}{\beta N} \log\left(\Lambda_1^N + \Lambda_2^N\right) \tag{7}$$

where $\Lambda_1 \geq \Lambda_2$ denote the two eigenvalues of the (positive) symmetric matrix $\boldsymbol{T}$. The free energy $f = \lim_{N \to \infty} f_N$ in the thermodynamic limit is

$$-\beta f = \log(\Lambda_1) = \log\left(e^K \cosh(h) + \sqrt{e^{2K} \sinh^2(h) + e^{-2K}}\right) \tag{8}$$

and hence is an analytic function of the magnetic field $H$ and the temperature $T$ for all real $H$ and positive $T$. The point $H = T = 0$ can be interpreted as sort of a critical point (in the sense that the correlation length becomes infinite), see e.g. (Baxter, 1982, pp. 32–38). The situation is generic for 1D models with short-ranged interactions which cannot have phase transitions at non-zero temperature (van Hove, 1950).

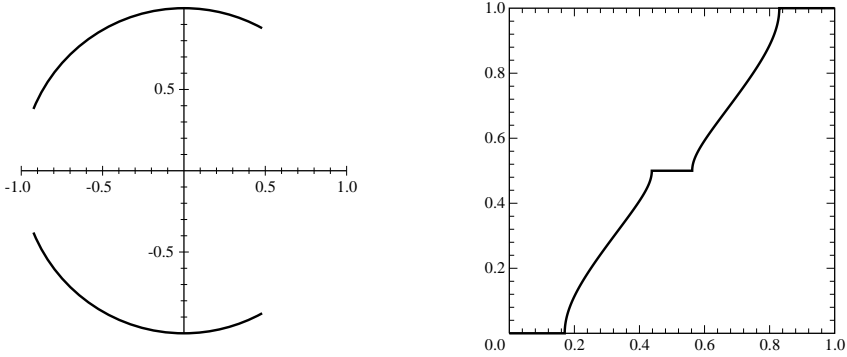

*Figure 2.* Magnetic field zeros of an alternating Ising chain and their integrated density

Before we move on to 2D systems, let us have a short look at the partition function $Z_N$ (5) for an Ising chain with uniform magnetic field $H_j = H$ and its zeros in the variable $\omega = \exp(2\beta H)$ (*Lee-Yang zeros*). According to the Lee-Yang theorem (Lee and Yang, 1952; Ruelle, 1969), for ferromagnetic coupling constants the zeros lie on the unit circle. For the simply periodic Ising chain ($J_j \equiv J$), the zeros in the thermodynamic limit fill a connected part of the unit circle densely (though not uniformly) leaving a gap around the intersection with the positive real axis which only closes for $T \to 0$ when the distribution of zeros becomes uniform. More generally, one obtains $p$ connected parts of zeros on the unit circle for periodic chains with unit cell of length $p$.

In Figure 2, we show the distribution of the magnetic field zeros on the unit circle and their integrated density (plotted against the angle in units of $2\pi$) for an *alternating* Ising chain ($p = 2$) with coupling constants $J_{2j-1} = J_a$ and $J_{2j} = J_b$, with numerical values $\exp(2\beta J_a) = 3/2$ and $\exp(2\beta J_b) = 3$ (i.e., $J_b/J_a \approx 2.71$). The derivation is given in the Appendix.

## 2.2.  SQUARE LATTICE ISING MODEL

Let us have a short look at the Ising model on the square lattice. In Figure 3, the Ising variables $\sigma_{j,k} \in \{1, -1\}$ are represented by black dots on the vertices of the square lattice, neighbouring pairs of spins interacting with coupling constants $K/\beta$ along horizontal and $L/\beta$ along vertical bonds, respectively. The energy for the Ising model in a magnetic field $H = h/\beta$ on a lattice of size $N \times M$ reads

$$\beta E(\boldsymbol{\sigma}) = -\sum_{j=1}^{N} \sum_{k=1}^{M} \left( K\sigma_{j,k}\sigma_{j+1,k} + L\sigma_{j,k}\sigma_{j,k+1} + h\sigma_{j,k} \right) \qquad (9)$$

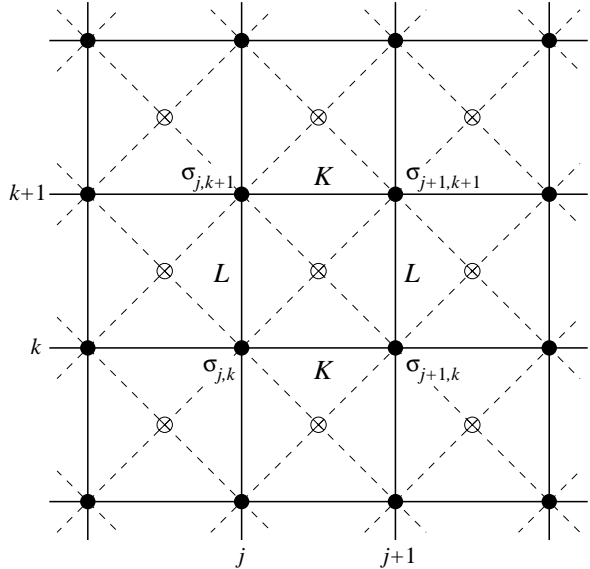

*Figure 3.*  The Ising model on the square lattice and its interpretation as an IRF model

where periodic boundary conditions in both horizontal and vertical directions are assumed (thus we actually work on a torus).

For the solvable zero-field case, there is a variety of approaches to calculate the free energy, see (Baxter, 1982) and references therein. Let us add how the model fits into the zoo of solvable lattice models in the sense of *commuting transfer matrices*. Figure 3 also shows another square lattice of dashed lines which contains additional vertices represented by open circles. Using this lattice, the Ising model can be reformulated as an IRF (*interaction-round-a-face*) model, where statistical weights are associated to elementary plaquettes (faces) rather than to bonds of the lattice. To do this, one defines a three-state model (with local states $\{1, 0, -1\}$, say) on the dashed lattice subject to the requirements that states on adjacent vertices of the lattice have to differ by one. In particular, the states 1 and $-1$ may not be neighbours on adjacent lattice sites (or, equivalently, the statistical weights of such configurations are zero). Then any "allowed" configuration has one sublattice filled with the state 0, while only 1 and $-1$ occur on the other sublattice. One can now find *face weights* for this model such that its partition function coincides with that of the Ising model, see e.g. (Wadati et al., 1989). For the zero-field case $H = 0$, these face weights can be parametrized by elliptic functions of a so-called *spectral parameter* such that they satisfy the YBE, which implies commutativity of transfer matrices for different spectral parameters. So, eigenvalues depend on the parameter but eigenvectors do not.

The physical quantity (often called *order parameter*) that distinguishes between the *ordered* and *disordered phases* of the *ferromagnetic* Ising model is the *magnetization*

$$m(H,T) = \lim_{N,M\to\infty} \frac{1}{NM} \sum_{j=1}^{N} \sum_{k=1}^{M} \langle \sigma_{j,k} \rangle = \lim_{N,M\to\infty} \langle \sigma_{j,k} \rangle \qquad (10)$$

which (for given coupling constants) is a function of the magnetic field $H$ and the temperature $T$. The second equality in Eq. (10) follows from *translational invariance*, which guarantees that the *local magnetization* $m_{i,j} = \lim_{N,M\to\infty} \langle \sigma_{j,k} \rangle$ is the same for all lattice points, while $\langle \sigma_{j,k} \rangle$ still depends on the system size. The magnetization can also be obtained as a derivative of the free energy $f(H,T)$ by $m(H,T) = -\partial f(H,T)/\partial H$. For sufficiently high temperature $(T > T_c)$, $m(H,T)$ is a smooth monotonously increasing function of the field, but at a certain temperature $T_c$ (the *critical* or *Curie temperature*), the derivative $\partial m(H,T_c)/\partial H$ (the *magnetic susceptibility*) becomes infinite at $H = 0$, and for $T < T_c$ the function $m(H,T)$ has a jump discontinuity at $H = 0$, see figure 1.1 in (Baxter, 1982). The jump corresponds to a *spontaneous magnetization*

$$m_0(T) = \lim_{H\to 0_+} m(H,T) = -\lim_{H\to 0_-} m(H,T) . \qquad (11)$$

The critical temperature $T_c$ is determined by

$$\sinh(2K)\sinh(2L) = 1 , \qquad H = 0 , \qquad (12)$$

which is just the condition that the Ising model is *self-dual*, i.e., invariant under *duality transformation* (Kramers and Wannier, 1941).

The spontaneous magnetization of the system vanishes in the disordered phase $(T > T_c)$ and is finite in the ordered phase $(T < T_c)$, behaving as $m_0(T) \sim (T_c - T)^{\beta}$ with the critical exponent[1] $\beta = 1/8$ as the critical point is approached $(T < T_c)$, see Figure 4. It shows also the *specific heat*

$$c_H(T) = -T \frac{\partial^2 f(H,T)}{\partial T^2} \qquad (13)$$

at constant magnetic field $H$, which displays a *logarithmic* divergence at the critical point (Onsager, 1944). Thus the corresponding critical exponent is $\alpha = 0$. Other critical exponents can be defined for the singular behaviour

---

[1] We stick to conventional notation since confusion between critical exponent $\beta$ and inverse temperature $\beta$ is unlikely.

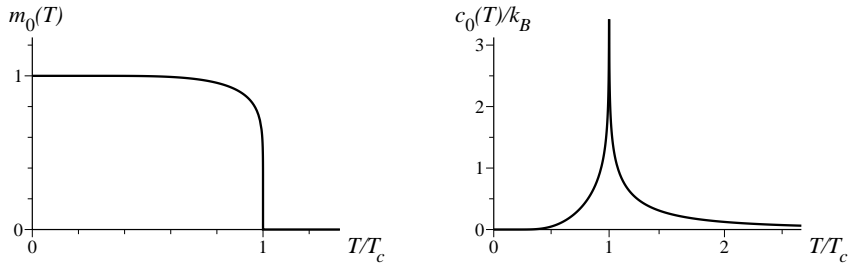

*Figure 4.* Spontaneous magnetization and specific heat of the isotropic Ising model

of physical quantities near critical points, see (Baxter, 1982) for details.

### 2.3. ISING QUANTUM CHAIN

The Ising quantum chain is defined by the Hamiltonian

$$\boldsymbol{H}_N = -\frac{1}{2}\left(\sum_{j=1}^{N}\varepsilon_j\,\sigma_j^x\sigma_{j+1}^x + \sum_{j=1}^{N}\sigma_j^z\right) \tag{14}$$

acting on the tensor product space $\bigotimes_{j=1}^{N}\mathbb{C}^2 \cong \mathbb{C}^{2^N}$. Here, $\sigma_j^x$ is defined as

$$\sigma_j^x = \underbrace{\mathbb{I}\otimes\mathbb{I}\otimes\ldots\otimes\mathbb{I}}_{j-1 \text{ factors}}\otimes\sigma^x\otimes\underbrace{\mathbb{I}\otimes\ldots\otimes\mathbb{I}\otimes\mathbb{I}}_{N-j \text{ factors}} \tag{15}$$

and $\sigma_j^z$ accordingly, where $\sigma^x$ and $\sigma^z$ denote Pauli matrices and $\mathbb{I}$ the identity, which in the usual basis have the form

$$\sigma^x = \begin{pmatrix} 0 & 1 \\ 1 & 0 \end{pmatrix}\;, \quad \sigma^z = \begin{pmatrix} 1 & 0 \\ 0 & -1 \end{pmatrix}\;, \quad \mathbb{I} = \begin{pmatrix} 1 & 0 \\ 0 & 1 \end{pmatrix}\;. \tag{16}$$

The Hamiltonian (14) can be viewed (see (Fradkin and Susskind, 1978; Kogut, 1979) and (Baxter, 1982, pp. 266-267)) as *anisotropic limit* of the transfer matrix of a *layered* zero-field square lattice Ising model at finite temperature, where the "horizontal" coupling term in Eq. (9) is replaced by $K_j\sigma_{j,k}\sigma_{j+1,k}$ (and $h = 0$), by performing the simultaneous limits $K_j \to 0$ and $L \to \infty$ while keeping $\varepsilon_j = K_j\exp(2L)$ finite. Therefore, the zero-temperature properties (i.e., eigenvalues and eigenvectors) of the quantum mechanical system defined by (14) correspond to the finite-temperature behaviour of the classical 2D Ising model. In particular, the critical line (12) of the square lattice Ising model corresponds to $\varepsilon_j \equiv 1$ in this limit.

The Ising quantum chain can be treated by similar methods as the 2D Ising model. The most popular technique following Lieb, Schultz and Mattis (1961) employs a Jordan-Wigner transformation to map the model onto a *free-fermion system*, and thereby reduces the problem of diagonalizing the Hamiltonian (14) (which is a $2^N \times 2^N$ matrix) to a diagonalization of a quadratic form (which amounts to diagonalizing an $N \times N$ matrix). From the set of its $N$ eigenvalues (*fermion energies* or *frequencies*), any eigenvalue of (14) can be obtained summing the elements in one of the $2^N$ different subsets.

To be more precise, one has to consider a modified Hamiltonian to avoid the appearance of non-local boundary terms in the fermionic language. It is given by the direct sum of the even part (under spin reversal) of the Hamiltonian (14) and the odd part of the same Hamiltonian with *antiperiodic* boundary conditions (i.e., $\sigma^x_{N+1} = -\sigma^x_1$). This is the *mixed sector* Hamiltonian , for details see (Baake et al., 1989).

## 2.4. CONFORMAL SYMMETRY

Two-dimensional lattice systems at criticality are closely related to conformal field theory (Belavin et al., 1984; Cardy, 1987; Ginsparg, 1990; Cardy, 1990; Christe and Henkel, 1993), for a more mathematical introduction to conformal field theory the reader is referred to (Gawędzki, 1989). This implies that the largest eigenvalues $\Lambda_1(N) \geq \Lambda_2(N) \geq \ldots$ of the transfer matrix are asymptotically degenerate at the critical point. The explicit behaviour for an isotropic system is (with $j > 1$)

$$\log\left(\frac{\Lambda_1}{\Lambda_j}\right) = \frac{2\pi}{N}x_j + o(N^{-1}) . \tag{17}$$

Conformal symmetry predicts the scaling exponents $x_j$ through irreducible representations of the *Virasoro algebra*

$$[L_m, L_n] = (m - n)L_{m+n} + \frac{c}{12}m(m^2 - 1)\delta_{m+n,0} , \quad m, n \in \mathbb{Z}. \tag{18}$$

Roughly speaking, $L_0$ is related to the logarithmic derivative of the transfer matrix (through a proper scaling limit) and the $L_m$ with negative integers $m$ act as ladder operators for the scaled gaps. Explicitly, one gets $x_j = \Delta + \bar{\Delta} + r + \bar{r}$ where $\Delta$ and $\bar{\Delta}$ are highest weights of two irreducible representations and $r, \bar{r} \in \mathbb{N}_0$. The generator $c$ in Eq. (18) commutes with all $L_n$ and is a complex number in any irreducible representation. It is called the *central charge* of the theory and plays a crucial rôle in a better understanding of the notion of universality. The value of the central charge is related (Blöte et al., 1986) to the finite-size corrections of the largest eigenvalue $\Lambda_1$ by

$$\log\left(\Lambda_1\right) = -\beta f N + \frac{\pi c}{6N} + o(N^{-1}) , \tag{19}$$

where $f$ denotes the bulk free energy per lattice site.

The critical Ising model is described by the simplest non-trivial *unitary minimal conformal field theory* with central charge $c = 1/2$. In this case, there are three irreducible representations which have highest weights $\Delta \in \{0, 1/2, 1/16\}$. The generating function (or character) of the corresponding modules (obtained as $\chi_\Delta(z) = \mathrm{tr}(z^{L_0})$) can be written (Baake, 1988) as

$$
\begin{aligned}
\chi_0(z) &= \sum_{n \in \mathbb{Z}} z^{4n^2+n}\, \Pi_V(z^2) &&= 1 + z^2 + z^3 + 2z^4 + 2z^5 + \ldots \\
\chi_{\frac{1}{2}}(z) &= \sum_{n \in \mathbb{Z}} z^{4n^2+3n+\frac{1}{2}}\, \Pi_V(z^2) &&= z^{\frac{1}{2}}\,(1 + z + z^2 + z^3 + 2z^4 + \ldots) \\
\chi_{\frac{1}{16}}(z) &= z^{\frac{1}{16}} \prod_{m=1}^{\infty} (1 + z^m) &&= z^{\frac{1}{16}}\,(1 + z + z^2 + 2z^3 + 2z^4 + \ldots)
\end{aligned}
\tag{20}
$$

where $\Pi_V(z)$ is the generating function of the number of partitions

$$
\Pi_V(z) = \prod_{m=1}^{\infty} \frac{1}{1 - z^m} \; .
\tag{21}
$$

The scaling (or universal) part of the Ising model partition function for periodic boundary conditions is a quadratic form in characters, and comprises the representations with conformal weights $(\Delta, \bar{\Delta}) = (0,0)$ (corresponding to the identity operator), $(\Delta, \bar{\Delta}) = (1/16, 1/16)$ (spin-density operator) and $(\Delta, \bar{\Delta}) = (1/2, 1/2)$ (energy-density operator).

For the Ising quantum chain (14), which alternatively can be viewed as the logarithmic derivative of the transfer matrix of the square lattice Ising model with respect to the spectral parameter, analogous relations to Eqs. (17) and (19) hold for the energy gaps $E_j - E_1$ and the finite-size corrections of the ground-state energy $E_1$, respectively. Here, the eigenvalues $E_j$ of the Hamiltonian play the part of $-\log(\Lambda_j)$. In addition, the scaling factor of the universal term has to be modified to correct for the anisotropy introduced by the Hamiltonian limit. For the mixed-sector Hamiltonian of the Ising quantum chain mentioned at the end of the previous subsection, the periodic sector contains weights $(\Delta, \bar{\Delta}) = (0,0)$, $(0, 1/2)$, $(1/2, 0)$, and $(1/2, 1/2)$, which gives the field theory of a free massless Majorana fermion.

## 3. Aperiodic Ising Chains

It is now time to turn to *aperiodic* Ising models, commencing with the simplest case of the classical 1D Ising chain (1). Here, we restrict ourselves to chains with a constant magnetic field $H_j = H$ and two *ferromagnetic* (i.e., positive) coupling constants $J_j \in \{J_a, J_b\}$, $j = 1, \ldots, N$. The actual sequence of coupling constants is determined by a *two-letter substitution* *rule* on the *alphabet* $\{a, b\}$. Substitution sequences are treated extensively in other parts of this volume (see also (Allouche and Mendès France, 1995)), thus we do not need to go into detail here.

A number of investigations has been performed on this or related models, see the references listed in the Bibliography. In particular, exact renormalization group transformations were used to study the ground-state and thermodynamic properties of Fibonacci Ising chains without and with a uniform magnetic field (Achiam et al., 1986), also for a mixture of ferro- and antiferromagnetic couplings (Tsunetsugu and Ueda, 1987). The zero-temperature ground-states for the case of uniform couplings and quasiperiodic field have also been investigated (Luck, 1987; Sire, 1993).

Of course, it is still impossible to observe a phase transition at non-zero temperature in these 1D systems, but nevertheless interesting properties arise. As an example, we consider the Lee-Yang zeros of the partition function in the field variable $\omega = \exp(2\beta H)$ and their distribution on the unit circle, compare (Baake et al., 1995; Simon et al., 1995) for complementary material.

As one of our main examples throughout this article we choose the *silver mean*[2] *substitution rule* $\varrho$

$$\varrho : \begin{array}{l} a \to b \\ b \to bab \end{array} , \qquad M_\varrho = \begin{pmatrix} 0 & 1 \\ 1 & 2 \end{pmatrix} , \qquad \lambda_\varrho^{\pm} = 1 \pm \sqrt{2} . \qquad (22)$$

Here, $M_\varrho$ is the corresponding *substitution matrix* with eigenvalues $\lambda_\varrho^{\pm}$. By iterated application of $\varrho$ on the initial word $w_0 = a$ one constructs words $w_{n+1} = \varrho(w_n) = w_n w_{n-1} w_n$ of increasing length

$$|w_n| = f_n + f_{n-1} , \qquad (23)$$

where the silver mean numbers are defined through

$$f_{n+1} = 2f_n + f_{n-1} , \qquad f_0 = 0 , \qquad f_1 = 1 . \qquad (24)$$

Two consecutive numbers are coprime, and $f_{n+1}/f_n \to \lambda_\varrho^+$ for $n \to \infty$.

To any finite word $w$ in the alphabet $\{a, b\}$ we associate an Ising chain (with periodic boundary conditions) with $N = |w|$ sites by choosing the coupling constant $J_j$ in Eq. (1) to be $J_a$ $(J_b)$ if the $j$th letter of $w$ is an $a$ $(b)$, respectively. Thus, we have two different local transfer matrices

$$\boldsymbol{T}_s = (\omega z_s)^{-1/2} \begin{pmatrix} \omega z_s & \omega^{1/2} \\ \omega^{1/2} & z_s \end{pmatrix} \qquad (25)$$

---

[2]The irrational number $\lambda_\varrho^+ = 1 + \sqrt{2} = [2; 2, 2, 2, \ldots]$ is called the *silver* mean for obvious reasons in comparison with the *golden* number $\tau = (1 + \sqrt{5})/2 = [1; 1, 1, 1, 1, \ldots]$.

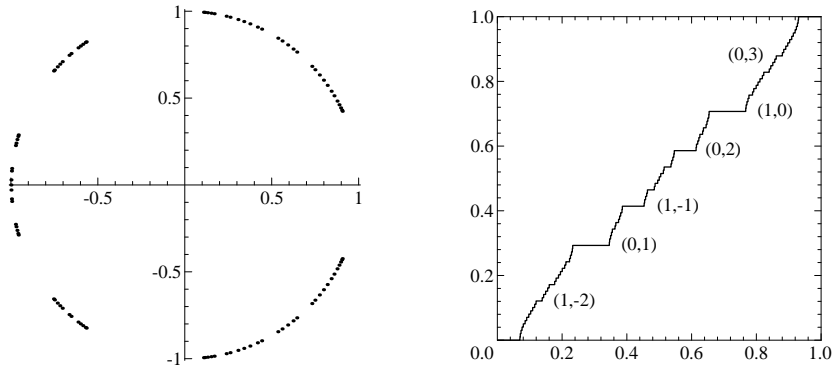

**Figure 5.** Lee-Yang zeros and their integrated density on the unit circle with gap labels

where $\omega = \exp(2\beta H)$, $z_s = \exp(2\beta J_s)$ and $s \in \{a, b\}$. This way, we obtain *periodic approximants* for the silver mean Ising chain. Denoting the corresponding product of local transfer matrices by $\mathcal{T}_n$, they satisfy the recurrence ($n > 0$)

$$\mathcal{T}_{n+1} = \mathcal{T}_n \mathcal{T}_{n-1} \mathcal{T}_n \qquad (26)$$

with $\mathcal{T}_0 = \mathbf{T}_a$ and $\mathcal{T}_1 = \mathbf{T}_b$. One can also derive a recursion relation for the partition function $\mathrm{tr}(\mathcal{T}_n)$ using the so-called *trace map*, see (Baake et al., 1993) for details. Note that the partition function is *real* for $|\omega| = 1$ (and real $z_s$) because $\mathbf{T}_s^* = \sigma^x \mathbf{T}_s \sigma^x$ with $\sigma^x$ given by Eq. (16).

From Eq. (25) it is obvious that the partition function is essentially a polynomial in the three variables $\omega$, $z_a$ and $z_b$. The quite general Lee-Yang theorem (Ruelle, 1969) guarantees (for ferromagnetic couplings, i.e., $z_s \geq 1$ for $s = a, b$) that all zeros in the complex variable $\omega$ lie on the unit circle $|\omega| = 1$ corresponding to imaginary values of the magnetic field. Figure 5 shows the zeros and their integrated density on the unit circle for the periodic approximant $w_6$ (of length $|w_6| = 99$) with couplings $z_a = 3/2$ and $z_b = 100$. In contrast to the periodic case (see above), the distribution in Figure 5 shows a characteristic gap structure reminiscent of the general picture for the electronic problem which is well understood in terms of the famous *gap labeling theorem* (Bellissard, 1986; Bellissard et al., 1992). Analogous results have also been observed for the Fibonacci (Baake et al., 1995) and Thue-Morse (Simon et al., 1995) Ising chains.

For the finite system attached to $w_n$, where the integrated density on the unit circle is approximated by the spectral step function, we obtain

plateaus at values

$$\frac{m}{f_n + f_{n-1}}, \qquad 0 \le m < f_n + f_{n-1}, \tag{27}$$

by construction. These are *independent* of the particular choice of the coupling constants. Replacing the *finite* system of length $p$ (with periodic boundary conditions) by an *infinite* system of period $p$, the integrated density becomes an absolutely continuous function with the plateaus (and the values on them) persisting. For length $p = 2$, this is explicitly shown in the Appendix; for the general statement, one employs the idea of Bloch's theorem.

Since $f_n$ and $f_{n-1}$ are coprime, one can write (Baake et al., 1993) $m$ in Eq. (27) as

$$m = \mu \cdot f_n + \nu \cdot f_{n-1} \tag{28}$$

with $\mu, \nu \in \mathbb{Z}$ (they can be made unique by proper restriction). Taking $(\mu, \nu)$ as label, we select a sequence of gaps in successive approximants which converge both in size and position. In our example of Fig. 5, the labels corresponding to the largest gaps are also given. In the limit $n \to \infty$ (which approaches our aperiodic chain) we obtain the dense set of gap values

$$\left\{ \frac{1}{\sqrt{2}} \left( \mu + \frac{\nu}{1 + \sqrt{2}} \right) \;\middle|\; \mu, \nu \in \mathbb{Z} \right\} \cap [0, 1]. \tag{29}$$

This coincides precisely with Bellissard's result for the electronic spectra of silver mean Hamiltonians. In the present context of magnetic chains, the meaning is that the derivative of the specific free energy, integrated along closed curves in the $\omega$-plane which do not hit the set of zeros, is quantized. More precisely,

$$\frac{1}{2\pi i} \oint_{\mathcal{C}} f'(\omega) \, d\omega \tag{30}$$

is a number of the set (29) if $\mathcal{C}$ is a path that starts at the origin, and returns to it through a gap on the unit circle.

Though we have sketched which gaps may appear, one still has to prove that they do not close in the limit. This can probably be shown with a slightly modified version of the constructive gap labeling of (Raymond, 1995) combined with the approach of the Appendix.

## 4. Aperiodic Ising Quantum Chains

In contrast to the 1D classical chain, the 2D Ising model has a phase transition at finite temperature. Consequently, as follows from the discussion of Section 2.3, the Ising quantum chain (14) inherits this phase transition with the temperature of the 2D model replaced by an expression in the coupling constants.

The Ising quantum chain and related models (notably the $XY$ quantum chain (Luck and Nieuwenhuizen, 1986; Satija and Doria, 1988)) with an aperiodic sequence of coupling constants have been studied frequently, a list of articles is given in the Bibliography. It also contains a number of papers devoted to the related problem of a 2D Ising model with *layered* aperiodicity. For both cases, most of the early results concentrate on the special examples of the Fibonacci (Doria and Satija, 1988; Iglói, 1988; Tracy, 1988; Benza, 1989; Ceccatto, 1989; Henkel and Patkós, 1992) and the Thue-Morse (Doria et al., 1989; Lin and Tao, 1990) Ising chains which very closely resemble the periodic situation. However, on the basis of two different examples of three-letter substitution rules, Tracy (1988) already conjectured that the critical properties coincide with those of the periodic model only if the corresponding substitution matrix has just one single eigenvalue of modulus greater than one. This has later been substantiated by investigation of models built on generalized Fibonacci sequences (Benza et al., 1990; You and Yang, 1990), where also the influence of particular choices of initial conditions of the substitution were noticed (Lin and Tao, 1992; You et al., 1992).

Finally, it was shown by scaling and perturbative arguments (Luck, 1993, 1994; Grimm and Baake, 1994) that the the critical behaviour depends on the *fluctuations* in the sequence of coupling constants, resulting in the same behaviour as the periodic case only if the fluctuations are *bounded*. In particular, this is the case for substitution rules for which all but the largest eigenvalue of the substitution matrix lie inside the unit circle. If the characteristic polynomial is irreducible over the integers, this is equivalent to the statement that the largest eigenvalue of the substitution matrix (the Perron-Frobenius eigenvalue, which is assumed to be larger than one) is a *Pisot-Vijayaraghavan number* or PV number, for short. These are real algebraic integers $\vartheta > 1$ with the property that all their algebraic conjugates (except $\vartheta$ itself) lie inside the unit circle.

As an example, we consider sequences defined by the substitution rules

$$\varrho_k : \begin{array}{l} a \to b \\ b \to ba^k \end{array} , \qquad M_k = \begin{pmatrix} 0 & k \\ 1 & 1 \end{pmatrix} , \qquad \lambda_k^\pm = \frac{1}{2} \left( 1 \pm \sqrt{4k+1} \right) . \quad (31)$$

For $k = 1$, this is nothing but the familiar *Fibonacci sequence* and $\lambda_1^+ = \tau = (1+\sqrt{5})/2 = [1; 1, 1, 1, 1, \ldots]$ is the *golden mean* which is a PV number. The second case ($k = 2$) is marginal, with eigenvalues $\lambda_2^+ = 2$ and $\lambda_2^- = -1$, whereas for $k \geq 3$ both eigenvalues $|\lambda_k^\pm| > 1$. As above, we define our sequences by iterated applications of the substitution rule $w_{n+1} = \varrho_k(w_n)$ on the initial word $w_0 = a$.

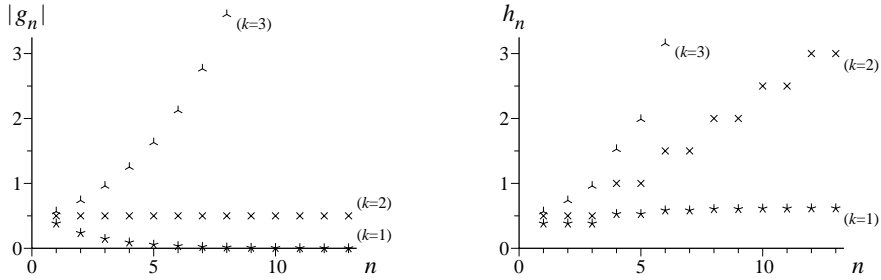

**Figure 6.** Fluctuations of letter frequencies in substitution sequences (31) for $k = 1, 2, 3$

The second eigenvalue $\lambda_k^-$ governs the behaviour of fluctuations in the sequence. To see this, consider the deviations from the mean frequency $p_a = 1 - 1/\lambda_k^+$ of the letter $a$

$$g(N) = |w_\infty(N)|_a - p_a N \; , \qquad g_n = g(|w_n|) \; , \qquad h_n = \max_{N \leq |w_n|} |g(N)| \; , \quad (32)$$

where $w_\infty(N)$ denotes the word obtained by truncating the limit word (fixed point of the substitution) $w_\infty$ after $N$ letters, and $|w|_a$ counts the number of occurrences of the letter $a$ in the word $w$. Generically, $|g_n|$ (and hence $h_n \geq |g_n|$) grows as the $n$th power of $\lambda_k^{(-)}$. In the marginal case, $|g_n|$ is bounded, but $h_n$ may still diverge logarithmically with the length of the word (i.e., linearly in $n$). In the PV case, we have *bounded fluctuations*, i.e., $|g_n| \to 0$ for $n \to \infty$ and $h_n$ is bounded (Bellissard et al., 1989). In Figure 6, this behaviour can clearly be seen for the substitution sequences (31) with $k = 1, 2, 3$.

If we consider the Ising quantum chain (14) with two different coupling constants $\varepsilon_a$ and $\varepsilon_b$ arranged according to the sequence of letters $a$ and $b$ in a substitution sequence, the condition for criticality reads ($p_b = 1 - p_a$)

$$p_a \log(\varepsilon_a) + p_b \log(\varepsilon_b) = 0 \qquad (33)$$

which can be solved by parametrizing the couplings by

$$\varepsilon_a = r^{-p_b} \; , \qquad \varepsilon_b = r^{p_a} \qquad (34)$$

in terms of an arbitrary positive real number $r$ (Benza, 1989). As mentioned before, the Hamiltonian (14) can be mapped onto a free fermion problem, which means that all its $2^N$ eigenvalues are obtained as the sums of $N$ fermion energies (*fermion frequencies*) with coefficients 0 or 1 (there are obviously $2^N$ different choices of the coefficients). Although we cannot present an analytic solution, this connection allows for a very efficient numerical treatment.

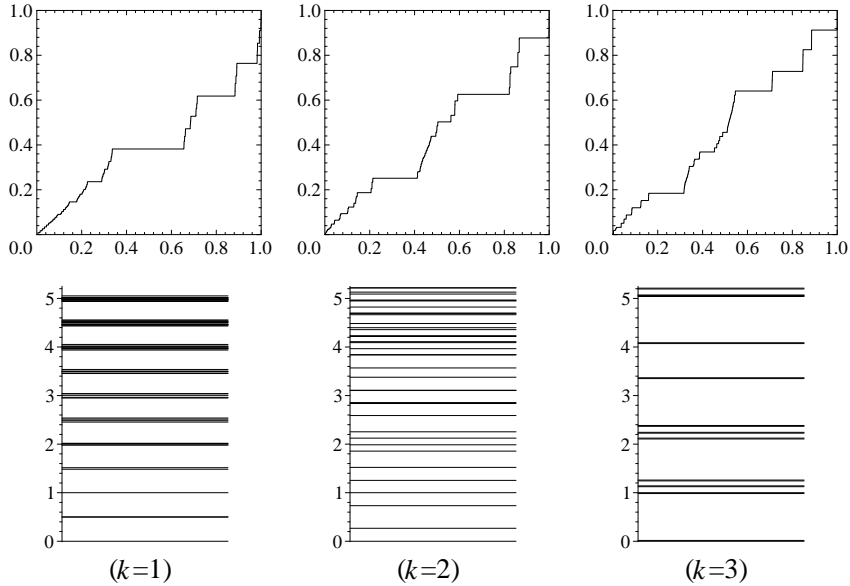

*Figure 7.* Integrated density of fermion frequencies and scaled spectrum

In Figure 7, the integrated densities of the critical fermion frequencies (normalized by their maximal values) are shown for the three substitution rules under study, for coupling constants given by Eq. (34) with $r = 3$. Here, we present the result for quantum chains obtained from $n = 12$ ($k = 1$), $n = 8$ ($k = 2$), and $n = 7$ ($k = 3$) iterations of the substitution rule, corresponding to chains of length $N = 233$, $N = 171$, and $N = 217$, respectively. Again, a characteristic gap structure similar to that observed in the Lee-Yang zeros of the 1D classical chains is apparent, and we suppose that the gap labeling theorem describes this situation as well. One important difference, however, occurs at the low-energy edge of the fermion spectrum. Compatibility with conformal invariance requires here a *linear* integrated density in the thermodynamic limit. Though this looks plausible in the case $k = 1$, the other integrated densities suffer from the fluctuations. To see this more clearly, one has to consider the appropriately scaled low-energy spectrum of the quantum chain.

In the second half of Figure 7, we show the low-lying part of the complete energy spectrum (of the mixed sector Hamiltonian mentioned above in Section 2.3), which has been scaled such that the third gap (i.e., the gap between the third excitation and the ground-state energy) has width one. For the Fibonacci case ($k = 1$), one clearly recognizes the equidistant scaled spectrum predicted by conformal invariance for the periodic system; a closer examination reveals that also the degrees of degeneracy (in the thermodynamic limit) match the prediction as described in Section 2.4. As an additional ingredient, one needs a modified scaling factor due to the anisotropy of the model. Here, this factor is known analytically and has the form $N/(2\pi v)$ where $v$ is the *fermion velocity* $v = 2\log(r)/(r - r^{-1})$ (Luck, 1993) with $r$ given in Eq. (34).

In the other two cases, the spectrum behaves quite differently from that of the Fibonacci chain. The normalization needed to keep the scaled gap finite grows faster than linearly in the system size $N$; for $k = 2$ it contains an additional logarithmic contribution in the system size, for $k = 3$ it grows as a power of $N$. This is in agreement with Luck's criterion (1993) for the relevance of aperiodicity — if it has *bounded* fluctuations, the aperiodicity is irrelevant and the critical behaviour is precisely the same as in the periodic case. However, if the fluctuations are *unbounded*, then the critical behaviour is indeed affected by the aperiodicity. The scaling behaviour in the marginal case with logarithmically diverging fluctuations ($k = 2$) has been investigated in (Berche et al., 1995) for the 2D model with layered aperiodicity.

## 5. Ising Model on Labyrinths

The most interesting case where one might hope to obtain analytic results is that of a 2D Ising model with an aperiodic distribution of coupling constants. As discussed above, the case of *layered* aperiodicity is covered by the Ising quantum chain, and numerous investigations (based on series expansion or numerical techniques such as Monte Carlo simulations) of Ising models on aperiodic graphs (mostly the Penrose tiling) have been performed, see the corresponding list in the Bibliography section. In contrast to the layered case, most results indicate that the critical behaviour remains unaffected by aperiodicity (Godrèche et al., 1986; Wilson and Vause, 1988; Sørensen et al., 1991). There are, however, specific influences of the local order on the Curie temperature (Bhattacharjee, 1994; Ledue et al., 1995) and on frustration in antiferromagnetic systems (Okabe and Niizeki, 1988; Oitmaa et al., 1990; Duneau et al., 1993; Ledue et al., 1993) while several hints on non-universal behaviour of Ising models on quasicrystals (Bhattacharjee et al., 1987; Minami and Suzuki, 1994) seem non-conclusive and need further investigation. First results for 3D Ising models (Bose, 1987; Okabe and Niizeki, 1990) show no signs of deviant critical properties due to aperiodicity.

The situation is similar for another class of critical phenomena comprising percolation of walks on quasiperiodic graphs. A number of publications devoted to this subject are listed in the Bibliography, see in particular the article by Briggs (1993) which contains numerical estimates for critical points and exponents of self-avoiding walks on several quasiperiodic graphs obtained by a series expansion analysis. Since walks are only sensitive to the topology of the graph, it is even harder to detect non-universal behaviour. This requires further investigations which go beyond the scope of this article.

Let us return to classical Ising models on 2D graphs. Apart from numerical approaches, the concept of "$Z$-invariance" (Baxter, 1978) has been applied to obtain analytic results for the eight-vertex and certain IRF models (Korepin, 1986, 1987; Choy, 1987, 1988; Antonov and Korepin, 1988) on (rhombic) duals of regular de Bruijn grids (de Bruijn, 1981). Here, we want to discuss the Ising model on 2D quasiperiodic graphs which are constructed from a rectangular grid. For definiteness, we concentrate on the so-called *Labyrinth tiling* (Sire et al., 1989) for which the underlying grid is built on silver mean sequences, compare (Baake et al., 1994). However, most results can easily be extended to more general classes of graphs.

### 5.1. THE LABYRINTH TILING

The silver mean chain is obtained by repeated application of the two-letter substitution rule $\varrho$ of Eq. (22) to the letter $w_0 = a$, i.e., $w_{n+1} = \varrho(w_n)$. From this, the Labyrinth tiling can be constructed by considering an orthogonal Cartesian product of two identical silver mean chains in proper geometric representation, i.e., the letters $b$ and $a$ are represented by long respectively short intervals with length ratio given by the silver mean $\lambda := \lambda_\varrho^+ = 1 + \sqrt{2}$. In this way, one arrives at a rectangular grid, and the Labyrinth tiling is obtained by choosing one of the two subgrids (defined as the set of vertices which can be reached from a chosen vertex by an even number of steps on the grid, or, in other words, as the two equivalence classes with respect to the corresponding equivalence relation) and connecting neighbouring points of the subgrid, see Figure 8. The graph defined that way is topologically equivalent to a square lattice.

Furthermore, we can naturally define a *dual graph* by connecting the points which lie on the other subgrid. Since the words $w_n$ obtained from the silver mean substitution rule (22) consist of an odd number of letters and are *palindromic*, the two dual graphs are equivalent to each other (even for a *finite* patch constructed from a word $w_n$) by reflection respectively by rotation through 90 degrees (which coincide due to the reflection symmetry of the graph with respect to the diagonal). This also implies a one-to-one map between the edges (bonds) of the graph and the dual graph; in fact, one has two choices due to the reflection symmetry of the Labyrinth. Either, one associates pairs of edges related by rotation through 90 degrees, or pairs that are mapped onto each other by reflection at a grid line. Here, we opt for the latter version because in this case the intersections between the Labyrinth and its dual (see Figure 9) occur between pairs of dual bonds, which is exactly what we need for the Ising model duality transformation discussed in Section 5.2 below.

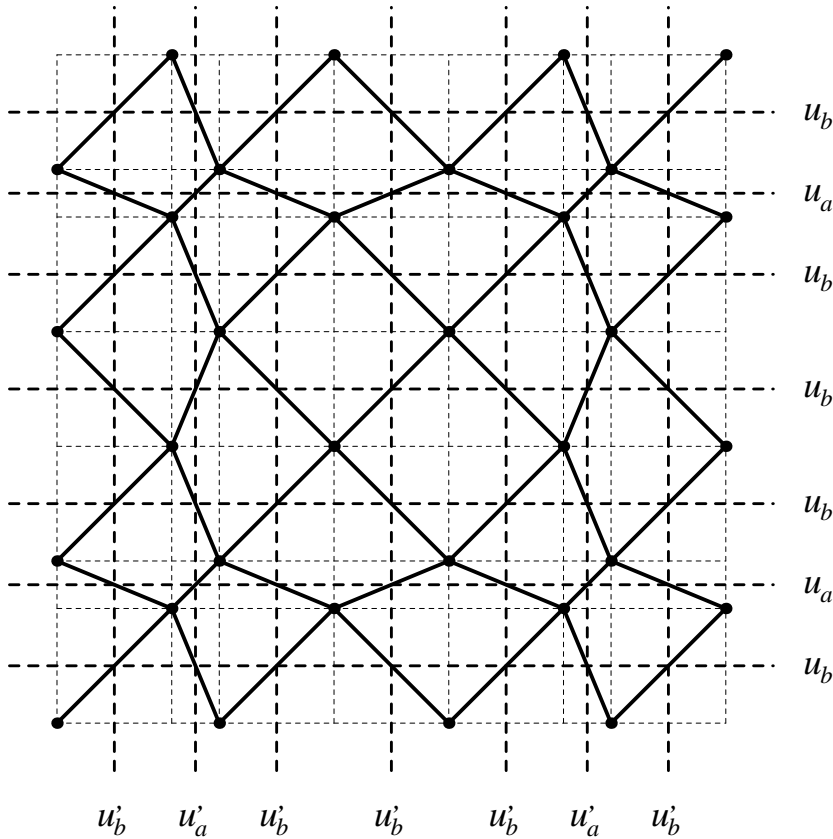

*Figure 8.* Labyrinth tiling with underlying grid and rapidity lines

Figure 9 shows the three different types of mutually dual tiles (dashed) and vertex configurations that appear in the Labyrinth. Their frequencies (in the infinite tiling) can easily be calculated from a induced substitution rule, which gives $5 - 2\lambda$ for the square, $6\lambda - 14$ for the kite and $10 - 4\lambda$ for the trapezoid, respectively. Since the tiling has fourfold rotational symmetry (Sire et al., 1989), the different orientations of the tiles (one for the square and four for the kite and the trapezoid) and corresponding vertex configurations occur equally often. The same is true of the three different edge types — long (denoted by $\ell$), medium ($m$) and short ($s$) — of the Labyrinth, their frequencies (normalized relative to the number of vertices, i.e., $\mu_\ell + \mu_m + \mu_s = 2$ as there are twice as many bonds as vertices) are calculated to give $\mu_\ell = 1$, $\mu_m = 2\lambda - 4$ and $\mu_s = 5 - 2\lambda$, respectively. Here, the long and short bonds occur in two, the bonds with medium length in four different orientations. Note that an analogous construction proceeding from a general two-letter substitution rule has four tile and vertex configurations (but still only three different edge lengths), one is missing here because $aa$ is not contained in the silver mean sequence as a subword.

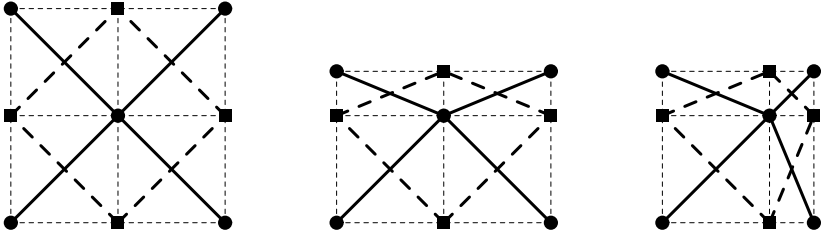

**Figure 9.** The mutually dual tiles and vertices of the Labyrinth

Now, we want to consider *periodic approximants* constructed from the finite patches described above. This is done using a periodic rectangular grid and performing the same construction as before. Since $|w_n| = 2|w_{n-1}| + |w_{n-2}|$ is odd, we cannot use the grid obtained from repeating $|w_n|$ periodically, because we need the existence of two distinct subgrids. Therefore, we instead *identify* the first and last letter in $|w_n|$ (which both are $b$, $n > 0$), or, in other words, construct our periodic grid by repeating the word $w'_n$ ($n > 0$) periodically (for both horizontal and vertical direction), where $w'_n$ denotes the word that is obtained by deleting the last letter in $w_n$. In this way, one arrives at a series of periodic approximants $\mathcal{L}_n$ with periods $|w_n| - 1$ which first of all do not contain any mismatches and furthermore are still equivalent to their duals $\mathcal{L}_n^*$ constructed from the other subgrid. This follows from the obvious fact that the two periodic sequences built on the word $w'_n$ and on its reverse are equivalent by translation.

## 5.2. ISING MODEL AND DUALITY TRANSFORMATION

Given a periodic approximant $\mathcal{L}_n$, we define our Ising model by placing the Ising variables on the vertices and associating coupling constants to the edges connecting them. Taking into account different orientations, we have eight different types of bonds, and therefore (in this setup) up to eight different coupling constants at our disposal. Let us, with $\beta = 1/k_B T$, use $K_{xy}/\beta$ and $L_{xy}/\beta$ ($xy \in \{aa, ab, ba, bb\}$) to denote the (ferromagnetic) coupling constants for bonds which are "raising" and "lowering" diagonals of a rectangle with abscissa $x$ and ordinate $y$ in the underlying grid, respectively, see Figure 10. In what follows, we consider the ferromagnetic zero-field case only, thus our parameter space is $\mathbb{R}_+^8$ since only the products of the coupling constants with the inverse temperature enter in the computation of the partition function. The reflection symmetry of the Labyrinth mentioned above implies that we can define a transformation on the space of coupling constants through

$$\begin{pmatrix} K_{xy} \\ L_{xy} \end{pmatrix} \longmapsto \begin{pmatrix} K_{yx} \\ L_{yx} \end{pmatrix} \tag{35}$$

which leaves the partition function of the Ising model invariant.

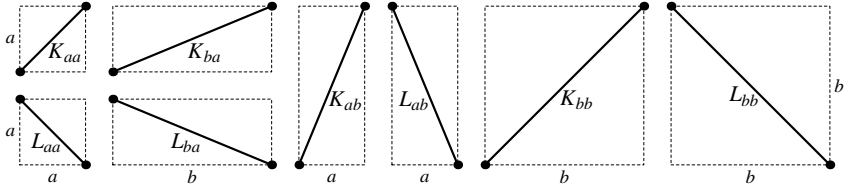

*Figure 10.* The eight different bonds and their associated coupling constants

The *duality transformation* $*$ is an analytic involution

$$* : \begin{pmatrix} K_{xy} \\ L_{xy} \end{pmatrix} \longmapsto \begin{pmatrix} K_{xy}^* \\ L_{xy}^* \end{pmatrix} \tag{36}$$

on $\mathbb{R}_+^8$, where $K_{xy}^* \in \mathbb{R}_+$ and $L_{xy}^* \in \mathbb{R}_+$ are defined by

$$\sinh(2K_{xy}^*)\sinh(2L_{xy}) = \sinh(2L_{xy}^*)\sinh(2K_{xy}) = 1 . \tag{37}$$

Under this transformation, the partition function is changed only trivially. This can be seen by investigating two exact graphical expansions of the partition function in terms of polygons on the Labyrinth (*high-temperature expansion*) respectively its dual (*low-temperature expansion*) which is again the Labyrinth, see (Baake et al., 1994) for details.

The Ising model is called *self-dual* if $K_{xy} = K_{xy}^*$ and $L_{xy} = L_{xy}^*$, i.e.,

$$S_{aa} = S_{ab} = S_{ba} = S_{bb} = 1 , \tag{38}$$

where $S_{xy}$ is defined as

$$S_{xy} := \sinh(2K_{xy})\sinh(2L_{xy}) . \tag{39}$$

Consider a smooth curve in parameter space along which all coupling constants are strictly increasing (ranging from 0 to infinity) and which is mapped onto itself by duality. It follows from Peierls' argument (Peierls, 1936; Ellis, 1985) that along such a curve one encounters at least one phase transition point. However, phase transition points correspond to parameter values where the free energy fails to be analytic, and thus must be mapped onto other phase transition points under duality. If one assumes that there is only *one* phase transition point on the curve (which is a natural assumption for the Ising model), it necessarily has to coincide with the intersection of the curve with the self-dual surface (which exists since the involution $*$ is analytic). Therefore, we suppose that Eq. (38) defines a four-dimensional critical surface in our eight-dimensional parameter space.

Although this result is much stronger than those obtained for Penrose tilings, where duality arguments together with correlation inequalities yield lower and upper bounds for the critical temperature (Bhattacharjee et al., 1987), it is not completely satisfactory. The reason is that, due to Peierls' argument, we expect a critical surface of codimension one dissecting the parameter space, or possibly an even more complicated scenario with several transitions, though this seems rather unlikely (because there appears to be no other natural order parameter besides the magnetization in our model, as long as we stay in the ferromagnetic regime). Still, the assumption of uniqueness of the phase transition is crucial in our argument — and we presently do not know of any rigorous argument to exclude the possibility of other transitions.

### 5.3. COMMUTING TRANSFER MATRICES

However, we can employ another approach based on commuting transfer matrices. As mentioned in Section 2.2, the Ising model on the square lattice can be reformulated as an IRF model, and the Boltzmann weights of the IRF model can be parametrized (in terms of elliptic functions of the spectral parameter) such that they satisfy the YBE, which implies commutativity of the corresponding row transfer matrices (for different values of the spectral parameter). Apart from the spectral parameter, the Boltzmann weights depend on one further parameter (the elliptic modulus) which basically plays the rôle of a temperature.

The values of the spectral parameter entering in the Boltzmann weights can be viewed as the differences of two so-called *rapidities*[3] associated to two sets of orthogonal lines (*rapidity lines*) which intersect in the corresponding face of the IRF model (respectively on the corresponding bonds of the Ising model). In fact, the values of these rapidities may be varied from one line to the other without destroying commutativity of row transfer matrices, because commutativity is a consequence of the *local* Yang-Baxter relation. This is the idea behind "Z-invariance" as introduced by Baxter (1978) for the eight-vertex model and later applied to a checkerboard Ising model (1986). The term "Z-invariance" stems from the fact that the YBE allows to "shift" the rapidity lines around without changing the partition function of the model (which is customarily denoted by the letter $Z$).

---

[3]This terminology originates in scattering theory in $1+1$ dimensions, where the YBE implies factorization of $S$-matrices and rapidities are used to parametrize the relativistic energy-momentum relation.

To make this more concrete, let us have a look at the example at hand. Besides the bonds of the Labyrinth and the underlying grid, Figure 8 also shows the rapidity lines (dashed) which form a dual rectangular grid. The horizontal (vertical) lines are labeled by variables $u_a$ and $u_b$ ($u'_a$ and $u'_b$) according to the type of edges of the underlying grid which they intersect orthogonally. Note that each pair of one horizontal and one vertical line intersects in one face of the grid respectively on one bond of the Labyrinth. Now, for arbitrary (in general complex) values of the four rapidities $u_a, u_b, u'_a, u'_b$, we can use the same parametrization for the Ising model coupling constants on the Labyrinth (respectively the corresponding Boltzmann weights for the IRF model on the underlying grid) as in the periodic or the checkerboard case (Baxter, 1986). Since only the differences of rapidities enter in the weights, and since we have the elliptic modulus as an additional parameter (which is *not* allowed to vary from one face to another), this yields a four-dimensional subspace of our eight-dimensional space of coupling constants where the model is solvable (in the sense of commuting transfer matrices).

The free energy in the thermodynamic limit can be calculated explicitly, it is given by the sum of free energies of the four periodic arrangements of the rapidities, properly weighted by their frequencies. Also, the *local spontaneous magnetization* can be computed, it is *independently* of the position in the lattice given by

$$\langle \sigma \rangle = \begin{cases} \left(1 - \Omega^{-2}\right)^{1/8} & \text{if } \Omega^2 > 1 \\ 0 & \text{if } \Omega^2 \leq 1 \end{cases} \tag{40}$$

where $\Omega$ is the elliptic modulus entering in the definition of the coupling constants. The local YBE finally reduces the calculation of the magnetization to that of the periodic case, where one can use Onsager's result. Unfortunately, the more general situation does not provide a new and independent method to calculate the magnetization. In conclusion, the critical points on the solvable surface are determined by $\Omega^2 = 1$, and the critical behaviour is exactly the same as in the periodic case, with a magnetic exponent $\beta = 1/8$.

The actual expression for the parametrization of the coupling constants is neither essential nor very illuminating, and the interested reader is referred to (Baxter, 1986; Baake et al., 1994) for the technical details of the calculation. However, to give an impression of the kind of constraints imposed by integrability, let us have a look at the corresponding conditions on the coupling constants. To this end, we introduce a third involution $\widehat{\phantom{a}}$,

$$\widehat{\phantom{a}} : \begin{pmatrix} K_{xy} \\ L_{xy} \end{pmatrix} \longrightarrow \begin{pmatrix} \widehat{K}_{xy} \\ \widehat{L}_{xy} \end{pmatrix} \tag{41}$$

on the space of coupling constants, where $\widehat{K}_{xy} \in \mathbb{R}_+$ and $\widehat{L}_{xy} \in \mathbb{R}_+$ are defined through

$$\sinh(2\widehat{K}_{xy}) \sinh(2K_{xy}) = \sinh(2\widehat{L}_{xy}) \sinh(2L_{xy}) = 1 . \tag{42}$$

Then the conditions on the couplings constants can neatly be summarized in the requirement that the equation

$$\Omega^2 \;\; = \;\; \prod_{j=1}^{4} \frac{\sinh(2M_j) \cosh(\widehat{M}_1 + \widehat{M}_2 + \widehat{M}_3 + \widehat{M}_4 - 2\widehat{M}_j)}{\sinh(2\widehat{M}_j) \cosh(M_1 + M_2 + M_3 + M_4 - 2M_j)} \tag{43}$$

has to be satisfied for *all* vertices of the Labyrinth, where $\Omega$ is an arbitrary constant (coinciding with the elliptic modulus used in the explicit parametrization) and where $M_j$ $(j = 1, 2, 3, 4)$ denote the four couplings around the particular vertex. Note that the value of $\Omega$ in Eq. (43) has to be the *same* for all vertices. For the Labyrinth, this results in altogether five relations between $\Omega$ and the eight couplings $K_{xy}, L_{xy}$. As can easily be verified, four of the five conditions can be simplified to

$$S_{aa} = S_{ab} = S_{ba} = S_{bb} = \Omega ; \tag{44}$$

the fifth, more complicated relation stems from the vertex with one long, one short and two medium length bonds. Therefore, the critical subspace $\Omega = 1$ on the solvable surface is in turn a subspace (of codimension one) of the self-dual surface defined by Eq. (38).

Eq. (43) clearly reveals the drastic constraints which integrability imposes on the coupling constants, and which is particularly apparent in the result for the local magnetization (40). For a general set of couplings, one expects that the magnetization at a particular vertex depends strongly on its local neighbourhood rather than giving the same result for *all* vertices of the graph. This behaviour is due to the fact that the value of $\Omega$ is the same for all vertices, which for instance makes it impossible to make one coupling around a vertex weaker without having to compensate this by enlarging one of the other coupling constants. This shows that the solvable cases are certainly not representative in general, and that one has to be careful if one wants to draw general conclusions from the analysis of the solvable subspace alone.

It should be clear from the discussion above that the method of commuting transfer matrices extends to *arbitrary* graphs which are constructed from a rectangular grid in an analogous way. Moreover, the result is always the same — all that matters are the relative frequencies of the different coupling constants. The critical behaviour remains the same as for the periodic Ising model, irrespective of the properties of the underlying sequence.

### 5.4. LOCAL MAGNETIZATION

From the last remark, it is clear that we need to gain more information about the critical behaviour of the generic case in order to understand the model completely. Unfortunately, one cannot in general compute the partition function analytically, and therefore we use numerical simulations as a first approach. The simplest quantity one can measure is the local magnetization at a particular vertex, and we expect to see that it will in general depend on the local neighbourhood of the vertex.

Here, we restrict ourselves to the case of three different coupling constants associated to the three different lengths of bonds, i.e., $K_{aa} = L_{aa} =: \beta J_s$, $K_{ab} = L_{ab} = K_{ba} = L_{ba} =: \beta J_m$, and $K_{bb} = L_{bb} =: \beta J_\ell$, where the subscripts $s$, $m$, and $\ell$ refer to short, medium, and long bonds, respectively. From Eq. (44), one now finds that the only solvable case among those is the periodic Ising model where all the coupling constants are equal.

For our numerical simulation, we employ the Swendsen-Wang Monte-Carlo algorithm (Binney et al., 1992) for the periodic approximant $\mathcal{L}_5$ constructed from the word $w_5$ of length $|w_5| = 41$ which is obtained after five applications of the silver mean substitution rule $\varrho$ of Eq. (22) to the initial word $w_0 = a$. For this patch of $(|w_5| - 1)^2 = 1600$ sites, we estimated the local magnetization at three different vertices, where we chose representatives of the three different types of vertex configurations, see Figure 9. The result is presented in Figure 11, where the local magnetization at the three vertices is plotted as a function of $T/T_c$. Figure 11(a) corresponds to the periodic case $J_s = J_m = J_\ell$, in Figure 11(b) the ratios of coupling constants are given by $J_s/J_m = 6/5$ and $J_\ell/J_m = 4/5$, and Figure 11(c) finally displays the results for $J_s/J_m = 7/5$ and $J_\ell/J_m = 3/5$. Here, the critical temperature $T_c$ was kept approximately constant by adjusting the coupling constants such that their average (per bond) is the same in all three cases, which is what we used to normalize the abscissa in Figure 11.

Clearly, the results show that the local magnetization becomes position-dependent once we leave the periodic case. That is exactly what one expects

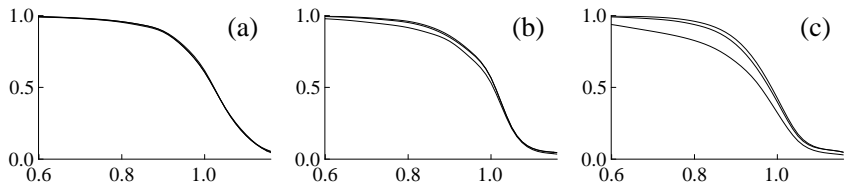

*Figure 11.* Local magnetization at three representative sites for different couplings

— by choosing the coupling along the long bonds to be smaller than the other two coupling constants one diminishes the local magnetization at the vertices surrounded by four long edges. On the other hand, Figure 11 does not show any indication that there might be more than one transition; the local magnetization at the three locations falls down at approximately the same value of the temperature. We believe that it is possible to prove the previous statement rigorously by applying suitable correlation inequalities (Griffiths, 1972); this will be investigated further.

## 6. Concluding Remarks

In this article, examples of the effects of aperiodicity on mathematical and physical properties of one- and two-dimensional Ising models have been discussed. Here, the aperiodicity we introduced is by no means random; in fact, the sequences we used to define our models (and the thermodynamic limit) are obtained from substitution rules and hence are completely deterministic.

One common feature that is apparent is the characteristic gap structure that shows up in various quantities for essentially one-dimensional situations. Obviously, the observed locations of the gaps obey the general gap labeling theorem for Schrödinger operators. Though we cannot present a rigorous proof for this statement at present, we believe that a constructive approach along the lines of (Raymond, 1995) is possible and hope to report on it soon.

As we see in the example of the Ising quantum chain, the critical behaviour of statistical systems may be changed by introducing aperiodicity. In this case, the fluctuations of the sequence of coupling constants determine whether the critical properties are affected or not. Of course, the same is true of a 2D Ising model with layered aperiodicity.

However, the situation appears to be different if one considers the 2D Ising model with an "isotropic" aperiodicity, for instance with a non-trivial rotational symmetry as in the case of the Labyrinth. Here, the available analytic and numerical results suggest that the critical behaviour remains unaffected for a large class of systems.

As stated in the beginning, it was not the purpose of this article to give final answers to the questions that arise in this context. However, we are confident that some of our observations can be made rigorous and help to improve the understanding of the mathematical and physical properties of non-periodic statistical systems.

### Acknowledgements

The material presented here is partially based on results we have obtained in interaction and cooperation with various other people. It is a pleasure to thank all of them, in particular Rodney Baxter, Bernard Nienhuis, Paul Pearce, Carmelo Pisani and Harald Simon. We also thank Robert Griffiths, Albertus Hof and Laurent Raymond for valuable discussions during the workshop. The authors would like to express their gratitude to the Fields Institute and the organizers of the workshop for their support. Financial support of Deutsche Forschungsgemeinschaft and Stichting voor Fundamenteel Onderzoek der Materie (Samenwerkingverband FOM/SMC Mathematische Fysica) is gratefully acknowledged.

### A. Magnetic field zeros of periodic Ising chains

Let us consider the partition function $Z_n = \text{tr}(\boldsymbol{T}^n)$ of a periodic model with a 2×2 transfer matrix $\boldsymbol{T}$. Assuming that $\det(\boldsymbol{T}) \neq 0$, we define $\varphi$ with $\Re(\varphi) \in [0, \pi]$ by

$$\cos(\varphi) = \frac{\text{tr}(\boldsymbol{T})}{2\sqrt{\det(\boldsymbol{T})}} \ . \tag{45}$$

By induction, one shows that

$$Z_n = 2\left[\det(\boldsymbol{T})\right]^{n/2} \cos(n\varphi) \tag{46}$$

and thus the equation $Z_n = 0$ is equivalent to $\cos(n\varphi) = 0$ which has $n$ real solutions $\varphi = (k - 1/2)\pi/n$ in the interval $[0, \pi]$. For $n \to \infty$, the solutions fill the complete interval $[0, \pi]$ with uniform density.

From this, one can easily calculate the density of partition function zeros in the variable $\omega = \exp(2\beta H)$ of the ferromagnetic periodic 1D Ising chain (i.e., $J_j \equiv J > 0$ and $H_j \equiv H$ in Eq. (1)) on the unit circle, which is done in some standard textbooks on statistical mechanics, see e.g. (Pathria, 1972, pp. 420–424). However, here we would like to treat the slightly more interesting example of an *alternating* Ising chain with ferromagnetic coupling constants $J_{2j-1} = J_a > 0$ and $J_{2j} = J_b > 0$ and uniform magnetic field $H_j = H$. Denoting the two corresponding local transfer matrices by $\boldsymbol{T}_a$ and $\boldsymbol{T}_b$, the partition function $Z_n$ of the *alternating chain* of length $N = 2n$ is

$$Z_n = \operatorname{tr}\left[(\boldsymbol{T}_a\boldsymbol{T}_b)^n\right] = \operatorname{tr}(\boldsymbol{T}_{ab}^n) \tag{47}$$

and hence the partition function of a *periodic* chain of length $n$ with transfer matrix $\boldsymbol{T}_{ab} = \boldsymbol{T}_a\boldsymbol{T}_b$. Introducing the notation $z_a = \exp(2\beta J_a) \geq 1$, $z_b = \exp(2\beta J_b) \geq 1$ and $\theta = -2i\beta H$, we find

$$\operatorname{tr}(\boldsymbol{T}_{ab}) = 2\,(z_a z_b)^{1/2}\cos(\theta) + 2\,(z_a z_b)^{-1/2} \tag{48}$$

$$\det(\boldsymbol{T}_{ab}) = z_a z_b + (z_a z_b)^{-1} - z_a z_b^{-1} - z_a^{-1} z_b \tag{49}$$

and therefore one can solve Eq. (45) for $\cos(\theta)$ to obtain

$$\cos\theta = \sqrt{(1 - z_a^{-2})(1 - z_b^{-2})}\,\cos(\varphi) - (z_a z_b)^{-1}\,. \tag{50}$$

The $N = 2n$ zeros of the partition function $Z_n$ are given by

$$\theta_k = \arccos\left[\sqrt{(1 - z_a^{-2})(1 - z_b^{-2})}\,\cos(\tfrac{(2k-1)\pi}{2n}) - (z_a z_b)^{-1}\right] \tag{51}$$

$(0 < \theta_k < \pi)$ and by their conjugate zeros $2\pi - \theta_k$, where $k = 1, 2, \ldots, n$. For $n \to \infty$, they form (if $z_a \neq z_b$ and $\beta$ is finite) *two* separated intervals with gap edges $\theta^\pm$ determined by inserting $\varphi = 0$ and $\varphi = \pi$ in Eq. (50). This yields

$$\theta^\pm = \arccos\left[\pm\sqrt{(1 - z_a^{-2})(1 - z_b^{-2})} - (z_a z_b)^{-1}\right] \tag{52}$$

$(0 < \theta^+ < \theta^- < \pi)$ and the corresponding conjugates $2\pi - \theta^\pm$. In contrast to the situation in the periodic Ising chain, the partition function zeros of the alternating chains have two gaps, one around $\theta = 0$ and the other in the vicinity of $\theta = \pi$. The latter closes if $z_a$ and $z_b$ become equal, the gap at $\theta = 0$ (which is related to the absence of a phase transition) only for $\beta \to \infty$, i.e., at zero temperature, where the distribution of zeros becomes uniform.

The integrated density of zeros $\mathcal{D}(\theta/2\pi)$ is defined as the limit of the ratio of the number of zeros $\theta_k < \theta$ and the total number of zeros, which is $2n$. Since the distribution of the values of $\varphi$ is uniform on $[0, \pi]$, this function has the following form

$$\mathcal{D}(\theta/2\pi) = \begin{cases} 0 & 0 \leq \theta \leq \theta^+ \\ \frac{1}{2\pi}\arccos\left[\frac{\cos(\theta) + (z_a z_b)^{-1}}{\sqrt{(1 - z_a^{-2})(1 - z_b^{-2})}}\right] & \theta^+ \leq \theta \leq \theta^- \\ \frac{1}{2} & \theta^- \leq \theta \leq \pi \\ 1 - \mathcal{D}(1 - \theta/2\pi) & \pi \leq \theta \leq 2\pi \end{cases} \tag{53}$$

and is shown in Figure 2. The corresponding root density is obtained by differentiation and diverges at the four gap edges $\pm\theta^\pm$ as $|\theta - (\pm\theta^\pm)|^{1/2}$ (*Lee-Yang edge singularity*).

Finally, let us mention that the same technique applies for periodic Ising chains with arbitrary period $p$, as long as the lengths of the chains $N = np$ are integer multiples of $p$. However, the right hand side of Eq. (45) is a polynomial of degree $p$ in $\cos(\theta/2)$ which generally makes it impossible to solve Eq. (45) for $\theta$ explicitly.

THE ISING MODEL AND ITS HISTORY

## CONFORMAL INVARIANCE

## APERIODIC ISING CHAIN

## APERIODIC QUANTUM SPIN CHAINS

## APERIODICALLY LAYERED 2D ISING MODEL

## Z-INVARIANT 2D LATTICE MODELS

## APERIODIC 2D LATTICE MODELS

---

[4]We warn the reader that the quoted page numbers occur *twice* in subsequent issues of the journal, the cited paper is contained in issue no. 14.

## PERCOLATION AND WALKS ON QUASIPERIODIC GRAPHS

## SUBSTITUTION SEQUENCES, TILINGS AND GAP LABELING